\documentclass[preprint,review,12pt]{elsarticle}

\usepackage{graphics}

\usepackage{amssymb}

\usepackage{subeqn}

\usepackage{bm}

\usepackage{multirow}

\usepackage{rotating}

\biboptions{sort&compress}

\journal{}

\begin{document}

\begin{frontmatter}



\title{The computation of strain rate tensor in multiple-relaxation-time lattice Boltzmann model}

\author[a]{Wenhuan Zhang\corref{cor1}}
\cortext[cor1]{Corresponding author, Tel and Fax: +86 574 87608744.}
\ead{zhangwenhuan@nbu.edu.cn}
\author[b]{Changsheng Huang}
\author[a]{Yihang Wang}
\author[c]{Baochang Shi}
\address[a]{Department of Mathematics and Ningbo Collaborative Innovation Center of Nonlinear Hazard System of Ocean and Atmosphere, Ningbo University, Ningbo 315211, P.R. China}
\address[b]{School of Mathematics and Physics, China University of Geosciences, Wuhan 430074, P.R. China}
\address[c]{School of Mathematics and Statistics, Huazhong University of Science and Technology, Wuhan 430074, P.R. China}

\begin{abstract}
    Multiple-relaxation-time (MRT) lattice Boltzmann (LB) model is an important class of LB model with lots of advantages over traditional
    single-relaxation-time (SRT) LB model. In addition, the computation of strain rate tensor is crucial in MRT-LB simulations of some complex flows. Up to now, there are only two formulas to compute the strain rate tensor in the MRT LB model.
    One is to compute the strain rate tensor by using non-equilibrium parts of macroscopic moments (Yu formula).
    The other is to compute the strain rate tensor by using non-equilibrium parts of density distribution functions (Chai formula).
    The mathematical expressions of these two formulas are so different that we do not know which formula to choose for computing the strain rate tensor in MRT LB model. In this paper, we study the relationship of these two formulas. It is found that Yu formula can be deduced from Chai formula in a particular procedure. However, these two formulas have their own advantages and disadvantages. Yu formula is more efficient in the computation aspect while Chai formula can be applied to more lattice patterns of MRT LB models. It is also found that deducing Yu formula for a particular lattice pattern from Chai formula is more convenient than the way proposed by Yu et al.
\end{abstract}

\begin{keyword}
lattice Boltzmann model\sep multiple relaxation time \sep strain rate tensor

\end{keyword}

\end{frontmatter}


\section{Introduction}
\label{1}The lattice Boltzmann equation (LBE), as a mesoscopic numerical method, has been widely used to simulate various complex fluid flows and has gained significant success because of its distinct advantages such as the natural parallelism of algorithm, simplicity of
programming and ease of dealing with complex boundary conditions \cite{ChenReview, Succi, AidunReview}.

Recently, the LBE method has also been used to the studies of blood flows\cite{6Boyd,10Krafczyk,11Artoli,12Rybicki}, non-Newtonian fluid flows\cite{5Gabbanelli,7Yoshino,8Vikhansky,Kruger1,9Chai,Huilgol,Papenkort,Conrad}, large eddy simulation of turbulent flows\cite{Yu,Malaspinasa_Sagaut1,Malaspinasa_Sagaut}, multiphase flows\cite{16Gross,Wang} and so on. In all the above studies, the computation of strain rate tensor is of key importance for simulations, therefore the computation of strain rate tensor has received increasing attention over the past decade \cite{Yu,9Chai,20Kruger,Chai2}.

For SRT LB model, it compute the strain rate tensor by using non-equilibrium parts of density distribution functions \cite{6Boyd,11Artoli,16Gross,20Kruger}. This particular computation method is local and has second-order accuracy in space, so it
is very suitable for the studies of flows in complex geometries and has great advantage over the finite difference method. Although the computation of strain rate tensor for SRT LB (also called LBGK) model has been studied a lot, the computation of strain rate tensor for MRT LB model is little investigated.

Yu et al. proposed to compute the strain rate tensor with non-equilibrium parts of macroscopic moments (Yu formula) for large eddy simulation (LES) of turbulent jets by using MRT LB model with nineteen discrete velocities in three dimensions (D3Q19) \cite{Yu}. Based on Yu formula, Premnath et al. developed the formula to compute the strain rate tensor with the external force effect for LES of turbulent flows \cite{Premnath_force}. Pattison et al. used the developed formula by Premnath et al. to compute the strain rate tensor for LES of turbulent flow in a straight square duct driven by a pressure gradient \cite{Pattison}. Some other works also used the formula developed by Yu et al. and Premnath et al. to compute the strain rate tensor for the LES of turbulent flows \cite{Jafari,Wu_Passive_scalar}.

More recently, Chai et al. proposed the other formula to compute the strain rate tensor with non-equilibrium parts of density distribution functions (Chai formula) for simulation of non-Newtonian fluid flows by using MRT LB model with nine discrete velocities in two dimensions (D2Q9) \cite{9Chai}. After this work, Chai formula was widely used to compute the strain rate tensor in many studies on non-Newtonian fluid flows by MRT-LB model \cite{Fallah,LiQX,ChenSG}.

To the knowledge of the authors, Yu formula and Chai formula are the only two formulas to compute the strain rate tensor in the MRT LB model up to now.
The computation formulas are so different that we are not sure which formula to choose for current simulations. In this paper, we firstly analyze the derivation of these two formulas. Then we study the characteristics of these two formulas.
Finally, we give our recommendation to the formula, which should be chosen for computing the strain rate tensor in the MRT LB model.

In the following, we firstly present He-Luo type MRT LB model with D3Q19 lattice \cite{dHumieres} as the starting point.
Based on the Chapman-Enskog analysis, we then deduce two formulas in computing the strain rate tensor in D3Q19 MRT LB model,
i.e., Yu formula and Chai formula. Thirdly, we study the relationship of above two formulas and analyze the advantages and
disadvantages of these two formulas. The suggestions to the choice of formulas when computing the strain rate tensor in MRT LB model are also given.
Finally, we give our conclusions.

\section{He-Luo D3Q19 MRT LB model}
In this paper, we compare the only two formulas (Yu formula and Chai formula) in computing the strain rate tensor in MRT LB model.
The macroscopic equilibrium moments of the MRT LB model are chosen to be derived from the equilibrium distribution functions of SRT LB (also called LBGK) model proposed by He and Luo \cite{He}, so we call this type of MRT LB model as He-Luo MRT LB model. In the following, we take He-Luo MRT LB model with D3Q19 lattice as an example to carry out the comparison.

The evolution equation of He-Luo D3Q19 MRT LB model is
\begin{equation}\label{mrteq}
  {f_\alpha }(\bm{x} + {\bm{c}_\alpha }{\delta _t},t + {\delta _t}) - {f_\alpha }(\bm{x},t) =  -
  \sum\limits_{i = 0}^{18} {{\Lambda _{\alpha i}}({f_i}(\bm{x},t) - f_i^{(eq)}(\bm{x},t))} {\rm{, }}\;\;\alpha
  {\rm{ = }}0-18,
\end{equation}
where $f_{l}(\bm{x},t)$ and $f_{l} ^{(eq)}(\bm{x},t)$ ($l=\alpha,i$) are the distribution function and equilibrium distribution function
of particles with velocity $\bm c_l$ at node $\bm{x}$ and time $t$, $\Lambda_{\alpha i}$ is the element located in $\alpha$ row and $i$ column
of $19\times19$ collision matrix $\bm \Lambda$. For He-Luo MRT model, the equilibrium distribution function is chosen as
\begin{equation}\label{edf}
f_i ^{(eq)} (\bm{x},t)= {\omega _i }\left[\rho+\rho_0\left({{{\bm{c}_i}\cdot \bm{u}} \over {c_s^2}} +
{{{{({\bm{c}_i}\cdot \bm{u})}^2}} \over {2c_s^4}} - {{|\bm{u}{|^2}} \over {2c_s^2}}\right)\right],
\end{equation}
where
\begin{equation}\label{omega}
{\omega _i } = \left\{ {\begin{array}{*{20}{l}}
{1/3,\;\;\;\;\;\; i  = 0,}\\
{1/18,\;\;\;\;\;\; i  = 1 - 6,}\\
{1/36,\;\;\;\; i  = 7 - 18,}
\end{array}} \right.
\end{equation}
$\bm{c}_i$ is defined as
 \begin{equation}
\footnotesize
 \begin{array}{c}
\{\emph{\textbf{c}}_{0}, \emph{\textbf{c}}_{1},\ldots,
\emph{\textbf{c}}_{18} \} =\nonumber
\\ \left\{
\begin{array}{ccccccccccccccccccc}
0  & 1 & -1 & 0 & 0 & 0 & 0 & 1 & -1 & 1 & -1 & 1 & -1 & 1 & -1 & 0 & 0 & 0 & 0 \\
0  & 0 & 0 & 1 & -1 & 0 & 0 & 1 &  1 & -1 & -1 & 0 & 0 & 0 & 0 & 1 & -1 & 1 & -1  \\
0  & 0 & 0 & 0 & 0 & 1 & -1 & 0 & 0 & 0 & 0 & 1 & 1 & -1 & -1 & 1&1 &-1 & -1
\end{array} \right\}c,
\end{array}
\label{HeLuoD3Q19}
\end{equation}
where $c=\delta_x/\delta_t$ is the particle velocity and $\delta_x$ and $\delta_t$ are the lattice spacing and time step, respectively. $c_s=c/\sqrt{3}$ is the sound speed. $\rho$ and $\bm{u}$ are the density and velocity of fluid. $\rho_0$ is the mean density.
The density fluctuation $\delta \rho=\rho-\rho_0$ is usually used instead of $\rho$ in the equilibrium distribution function to reduce
the numerical effects due to the round-off error \cite{dHumieres,Skordos}, i.e., the following equilibrium distribution function is commonly used,
\begin{equation}\label{edf}
f_i ^{(eq)} (\bm{x},t)= {\omega _i }\left[\delta \rho+\rho_0\left({{{\bm{c}_i}\cdot \bm{u}} \over {c_s^2}} +
{{{{({\bm{c}_i}\cdot \bm{u})}^2}} \over {2c_s^4}} - {{|\bm{u}{|^2}} \over {2c_s^2}}\right)\right].
\end{equation}
In the following, we suppose $c=1$ such that the relevant quantities are dimensionless.

The above equation can also be written in a vector form:
\begin{equation}\label{mrteqv}
|f(\bm{x} + {\bm{c}_\alpha }{\delta _t},t + {\delta _t})\rangle - |f(\bm{x},t)\rangle =  - \bm \Lambda
(|f(\bm{x},t)\rangle - |{f^{(eq)}}(\bm{x},t))\rangle,
\end{equation}
where $|f(\bm{x},t)\rangle=(f_0(\bm{x},t),f_1(\bm{x},t),\cdots,f_{18}(\bm{x},t))^{'}$ is a column vector, $|f^{(eq)}(\bm{x},t)\rangle$ and $|f(\bm{x} + {\bm{c}_\alpha }{\delta _t},t + {\delta _t})\rangle$ have similar definitions and the
superscript $'$ represents the transpose operator. For D3Q19 MRT model, a $19\times19$ transformation matrix is defined as \cite{dHumieres}:
{\tiny
\begin{equation}
\mathbf{T}=\left( {\begin{array}{*{20}c}
   1   &  1 & 1 & 1 & 1 & 1 & 1 & 1 & 1 & 1 & 1 & 1 & 1 & 1 & 1 & 1 & 1 & 1 & 1  \\
   -30 & { - 11} & { - 11} & { - 11} & { - 11} & { - 11} & { - 11} & 8 & 8 & 8 & 8 & 8 & 8 & 8 & 8 & 8 & 8 & 8 & 8  \\
   12  & { - 4} & { - 4} & { - 4} & { - 4} & { - 4} & { - 4} & 1 & 1 & 1 & 1 & 1 & 1 & 1 & 1 & 1 & 1 & 1 & 1  \\
   0  & 1 & { - 1} & 0 & 0 & 0 & 0 & 1 & { - 1} & 1 & { - 1} & 1 & { - 1} & 1 & { - 1} & 0 & 0 & 0 & 0  \\
   0  & { - 4} & 4 & 0 & 0 & 0 & 0 & 1 & { - 1} & 1 & { - 1} & 1 & { - 1} & 1 & { - 1} & 0 & 0 & 0 & 0  \\
   0 & 0 & 0 & 1 & { - 1} & 0 & 0 & 1 & 1 & { - 1} & { - 1} & 0 & 0 & 0 & 0 & 1 & { - 1} & 1 & { - 1}  \\
   0 & 0 & 0 & { - 4} & 4 & 0 & 0 & 1 & 1 & { - 1} & { - 1} & 0 & 0 & 0 & 0 & 1 & { - 1} & 1 & { - 1}  \\
   0 & 0 & 0 & 0 & 0 & 1 & { - 1} & 0 & 0 & 0 & 0 & 1 & 1 & { - 1} & { - 1} & 1 & 1 & { - 1} & { - 1}  \\
   0 & 0 & 0 & 0 & 0 & { - 4} & 4 & 0 & 0 & 0 & 0 & 1 & 1 & { - 1} & { - 1} & 1 & 1 & { - 1} & { - 1}  \\
   0 & 2 & 2 & { - 1} & { - 1} & { - 1} & { - 1} & 1 & 1 & 1 & 1 & 1 & 1 & 1 & 1 & { - 2} & { - 2} & { - 2} & { - 2}  \\
   0 & { - 4} & { - 4} & 2 & 2 & 2 & 2 & 1 & 1 & 1 & 1 & 1 & 1 & 1 & 1 & { - 2} & { - 2} & { - 2} & { - 2}  \\
   0 & 0 & 0 & 1 & 1 & { - 1} & { - 1} & 1 & 1 & 1 & 1 & { - 1} & { - 1} & { - 1} & { - 1} & 0 & 0 & 0 & 0  \\
   0 & 0 & 0 & { - 2} & { - 2} & 2 & 2 & 1 & 1 & 1 & 1 & { - 1} & { - 1} & { - 1} & { - 1} & 0 & 0 & 0 & 0  \\
   0 & 0 & 0 & 0 & 0 & 0 & 0 & 1 & { - 1} & { - 1} & 1 & 0 & 0 & 0 & 0 & 0 & 0 & 0 & 0  \\
   0 & 0 & 0 & 0 & 0 & 0 & 0 & 0 & 0 & 0 & 0 & 0 & 0 & 0 & 0 & 1 & { - 1} & { - 1} & 1  \\
   0 & 0 & 0 & 0 & 0 & 0 & 0 & 0 & 0 & 0 & 0 & 1 & { - 1} & { - 1} & 1 & 0 & 0 & 0 & 0  \\
   0 & 0 & 0 & 0 & 0 & 0 & 0 & 1 & { - 1} & 1 & { - 1} & { - 1} & 1 & { - 1} & 1 & 0 & 0 & 0 & 0  \\
   0 & 0 & 0 & 0 & 0 & 0 & 0 & { - 1} & { - 1} & 1 & 1 & 0 & 0 & 0 & 0 & 1 & { - 1} & 1 & { - 1}  \\
   0 & 0 & 0 & 0 & 0 & 0 & 0 & 0 & 0 & 0 & 0 & 1 & 1 & { - 1} & { - 1} & { - 1} & { - 1} & 1 & 1  \\
 \end{array} } \right),
\end{equation}
}which can transform the distribution function into the moment with the linear mapping $\bm m=\mathbf{T}{|f\rangle}$ and $\bm m^{(eq)}=\mathbf{T}{|f^{(eq)}\rangle}$,
and simultaneously convert the collision matrix into a diagonal one by $\bm {\hat  \Lambda} = \mathbf{T}\bm \Lambda \mathbf{T}^{-1}$. Thus, we can further write Eq. (\ref{mrteqv}) as
\begin{equation}\label{mrteqvT}
|f({\bm{x}} + {{\bm{c}}_\alpha }{\delta _t},t + {\delta _t})\rangle  - |f({\bm{x}},t)\rangle  =  - {\mathbf{T}^{ - 1}} \bm {\hat  \Lambda}(\bm m(\bm x,t) - {\bm m^{(eq)}}(\bm x,t)),
\end{equation} where $\bm m$ is defined as
 \begin{equation}\label{moment}
 \scriptsize
\bm m = (\delta \rho,e,\epsilon,\rho_0 u_x,q_x,\rho_0 u_y,q_y,\rho_0 u_z,q_z,3p_{xx},3\pi_{xx},p_{\omega \omega},\pi_{\omega \omega},p_{xy},p_{yz},p_{xz},t_{x},t_{y},t_{z})',
\end{equation}
and $\bm m^{(eq)}$ is written as
{\scriptsize
\begin{equation}\label{moment_eq}
\bm m^{(eq)} = (\delta \rho,e^{(eq)},\epsilon^{(eq)},\rho_0 u_x,q^{(eq)}_x,\rho_0 u_y,q^{(eq)}_y,\rho_0 u_z,q^{(eq)}_z,3p^{(eq)}_{xx},3\pi^{(eq)}_{xx},p^{(eq)}_{\omega \omega},\pi^{(eq)}_{\omega \omega},p^{(eq)}_{xy},p^{(eq)}_{yz},p^{(eq)}_{xz},t^{(eq)}_{x},t^{(eq)}_{y},t^{(eq)}_{z})'.
\end{equation}} where the equilibrium moments are
\begin{subequations}
\label{meq}
\footnotesize
\begin{equation}
\left.
\begin{array}{c}
    e^{(eq)}=-11\delta \rho+\frac{19}{\rho_0}\bm{j}\cdot \bm{j}, \\
   \epsilon^{(eq)}=\omega_{\epsilon}\delta \rho+\frac{\omega_{\epsilon j}}{\rho_0}\bm{j}\cdot \bm{j},
\end{array}
\right \}
\end{equation}
\begin{equation}
\left.
\begin{array}{c}
  q_x^{(eq)}=-2j_x/3, \\
  q_y^{(eq)}=-2j_y/3, \\
  q_z^{(eq)}=-2j_z/3, \\
\end{array}
\right \}
\end{equation}
\begin{equation}
\left.
\begin{array}{c}
  3p_{xx}^{(eq)}=\frac{1}{\rho_0}(3j_x^2-\bm{j}\cdot \bm{j}), \\
  p_{\omega \omega}^{(eq)}=3\omega_{xx}p_{xx}^{(eq)}, \\
\end{array}
\right \}
\end{equation}
\begin{equation}
\left.
\begin{array}{c}
  3\pi_{xx}^{(eq)}=\frac{1}{\rho_0}(j_y^2-j_z^2), \\
  \pi_{\omega \omega}^{(eq)}=3\omega_{xx}\pi_{xx}^{(eq)}, \\
\end{array}
\right \}
\end{equation}
\begin{equation}
\left.
\begin{array}{c}
  p_{xy}^{(eq)}=\frac{1}{\rho_0} j_x j_y, \\
  p_{yz}^{(eq)}=\frac{1}{\rho_0} j_y j_z, \\
  p_{xz}^{(eq)}=\frac{1}{\rho_0} j_x j_z, \\
\end{array}
\right \}
\end{equation}
\begin{equation}
\left.
\begin{array}{c}
  t_{x}^{(eq)}=0, \\
  t_{y}^{(eq)}=0, \\
  t_{z}^{(eq)}=0, \\
\end{array}
\right \}
\end{equation}
\end{subequations}
where $\bm{j}=(\rho_0u_x,\rho_0u_y,\rho_0u_z)'$, $j_i=\rho_0u_i\;(i=x,y,z)$, $\omega_\epsilon=3$, $\omega_{\epsilon j}=-11/2$ and $\omega_{xx}=-1/2$. It should be noted that $\omega_\epsilon$, $\omega_{\epsilon j}$ and $\omega_{xx}$ do not have much effect on the recovered Navier-Stokes equation. To attain an optimized stability of the model, we can adjust these parameters through linear analysis \cite{dHumieres}. In the following, we can see from Yu formula that these parameters do not have effect on the computation of strain rate tensor either.



The diagonal collision matrix $\bm{\hat \Lambda}$ is
\begin{equation}\label{sD3Q15}
\bm {\hat  \Lambda}\equiv diag(s_0,s_1,s_{2}, s_3, s_{4}, s_5, s_{6},s_7, s_{8},
s_{9}, s_{10}, s_{11}, s_{12}, s_{13}, s_{14},s_{15}, s_{16}, s_{17}, s_{18}),
\end{equation}
where $s_0$, $s_3$, $s_5$ and $s_7$ are the relaxation parameters corresponding to the conserved moments. The values of these parameters do not affect the recovered macroscopic N-S equations, which are always set to be zeros. In addition, Eq. (\ref{sD3Q15}) can also be written as
\begin{equation}\label{sD3Q15meaning}
\bm {\hat  \Lambda}\equiv diag(s_c,s_{e}, s_{\epsilon},s_c, s_{q}, s_c, s_{q}, s_c, s_{q},
s_{\nu},s_{\pi}, s_{\nu}, s_{\pi},s_{\nu}, s_{\nu}, s_{\nu}, s_{t},s_{t},s_{t}),
\end{equation}
where $s_c$, $s_e$, $s_\epsilon$, $s_q$, $s_\nu$, $s_\pi$ and $s_t$ are the parameters corresponding to the conserved moments, the moments related to kinetic energy, energy square, energy flux, strain rate tensor, et al. He-Luo D3Q19 MRT LB model can recovered to the incompressible Navier-Stokes equation by the Chapman-Enskog analysis. In the following, the discussion is based on this model.

\section{Two formulas in computing the strain rate tensor in MRT LB model}
Yu et al. proposed one formula to compute the strain rate tensor $S_{\alpha \beta} = (\partial_\alpha u_\beta + \partial_\beta u_\alpha)/2$ for He-Luo D3Q19 MRT LB model \cite{Yu}. In this formula, the strain rate tensor is computed from non-equilibrium parts of moments. The computation formula is
\begin{subequations}
\label{LuoS}
\begin{equation}
{{S}_{xx}}\approx -\frac{\varepsilon}{38{{\rho }_{0}}{{\delta }_{t}}}({{s}_{1}}m_{1}^{(1)}+19{{s}_{9}}m_{9}^{(1)}),
\end{equation}
\begin{equation}
{{S}_{yy,zz}}\approx -\frac{\varepsilon}{76{{\rho }_{0}}{{\delta }_{t}}}\left[ 2{{s}_{1}}m_{1}^{(1)}-19{{s}_{9}}(m_{9}^{(1)}\mp 3m_{11}^{(1)}) \right],
\end{equation}
\begin{equation}
{{S}_{xy,yz,zx}}\approx -\frac{3\varepsilon{{s}_{9}}}{2{{\rho }_{0}}{{\delta }_{t}}}m_{13,14,15}^{(1)},
\end{equation}
\end{subequations}
where $\varepsilon m_i^{(1)} \approx m_i-m_i^{(eq)}$ is the non-equilibrium part of moment $m_i$, $m_i$ and $m_i^{(eq)}$ are the $i$-th element of vector $\bm m$ and $\bm m^{(eq)}$.

Chai et al. proposed the other formula to compute the strain rate tensor for D2Q9 MRT LB model \cite{9Chai,Chai2}. This formula can also be used for D3Q19 MRT LB model. In this formula, the strain rate tensor is computed from non-equilibrium parts of density distribution functions. The computation formula is
\begin{equation}
\label{ChaiS}
\bm S=-\frac{\sum\nolimits_{i}{{\bm{c}_{i}}{\bm{c}_{i}}{{\left( {\mathbf{T}^{-1}}\bm{{\hat{\Lambda }}}\mathbf{T} \right)}_{ij}}f_{j}^{(neq)}}}{2\rho_0 c_{s}^{2}{{\delta }_{t}}},
\end{equation}
where $\bm S$ is strain rate tensor, $f_{j}^{(neq)}={{f}_{j}}-f_{j}^{(eq)}$. If $\bm{{\hat{\Lambda }}}=\frac{1}{\tau} \mathbf{I}$, where $\tau$ is the relaxation time in LBGK model and $\mathbf{I}$ is the unit matrix, the above expression becomes
\begin{equation}
\label{ChaiS_srt}
\bm S =  - \frac{{\sum\nolimits_i {{\bm c_i}{\bm c_i}\left( {{f_i} - f_i^{(eq)}} \right)} }}{{2{\rho _0}c_s^2{\delta _t}\tau }},
\end{equation}
which is exactly the strain rate tensor computation formula for LBGK model. Therefore, Chai formula is an extension of strain rate tensor computational formula for LBGK model.

\section{Derivation of two formulas in computing the strain rate tensor in MRT LB model}
\subsection{Derivation of Yu formula in computing the strain rate tensor in MRT LB model}
\label{subsec-Yu-derivation}
We first expand the density distribution function and the derivatives of time and space as
\begin{subequations}
\label{ftxeps}
\begin{equation}
{{f}_{i}}=f_{i}^{(0)}+\varepsilon f_{i}^{(1)}+{{\varepsilon }^{2}}f_{i}^{(2)}+\cdots
\end{equation}
\begin{equation}
\label{teps}
{{\partial }_{t}}=\varepsilon {{\partial }_{{{t}_{1}}}}+{{\varepsilon }^{2}}{{\partial }_{{{t}_{2}}}}
\end{equation}
\begin{equation}
\label{xeps}
{{\partial }_{\alpha}}=\varepsilon {{\partial }_{\alpha1}}.
\end{equation}
\end{subequations}
Substituting the above expansions into Eq. (\ref{mrteq}), we obtain the zero-, first-, and second-order equations in $\varepsilon$,
\begin{subequations}
\label{difepsorder}
\begin{equation}
\label{feps0order}
{{\varepsilon }^{0}}:\ f_{i}^{(0)}=f_{i}^{(eq)},
\end{equation}
\begin{equation}
\label{feps1order}
{{\varepsilon }^{1}}:\ {{D}_{1i}}f_{i}^{(0)}=-\frac{1}{{{\delta }_{t}}}{{\Lambda }_{ij}}f_{j}^{(1)},
\end{equation}
\begin{equation}
{{\varepsilon }^{2}}:\ {{\partial }_{{{t}_{2}}}}f_{i}^{(0)}+{{D}_{1i}}f_{i}^{(1)}+\frac{{{\delta }_{t}}}{2}D_{1i}^{2}f_{i}^{(0)}=-\frac{1}{{{\delta }_{t}}}{{\Lambda }_{ij}}f_{j}^{(2)},
\end{equation}
\end{subequations}
where ${{D}_{1i}}={{\partial }_{{{t}_{1}}}}+{{c}_{i\gamma}}{{\partial }_{\gamma1}}$.

If we rewrite Eq. (\ref{difepsorder}) in the vector form and multiply the matrix $\mathbf{T}$ on both sides of them, the corresponding equations in the moment space can be derived
\begin{subequations}
\label{difepsorderm}
\begin{equation}
{{\varepsilon }^{0}}:\ {\bm{m}^{(0)}}={\bm{m}^{(eq)}},
\end{equation}
\begin{equation}
\label{difeps1orderm}
{{\varepsilon }^{1}}: {{{\tilde\mathbf{D}}}_{1}}{\bm{m}^{(0)}}=-{\hat\mathbf{\Lambda }}'{\bm{m}^{(1)}},
\end{equation}
\begin{equation}
{{\varepsilon }^{2}}:\ {{\partial }_{{{t}_{2}}}}{\bm{m}^{(0)}}+{{{\tilde\mathbf{D}}}_{1}}(\mathbf{I}-\frac{{\hat{\mathbf{\Lambda }}}'}{2}){\bm{m}^{(1)}}=-{\hat\mathbf{\Lambda }}'{\bm{m}^{(2)}},
\end{equation}
\end{subequations}
where ${\hat{\mathbf{\Lambda} }}'=\hat{\mathbf{\Lambda} }/{{\delta }_{t}}$, $\hat\mathbf{\Lambda }={\mathbf{T}}{\mathbf{\Lambda}}\mathbf{T}^{-1}$, ${{{\tilde{\mathbf{D}}}}_{1}}=\mathbf{T}{{\mathbf{D}}_{1}}{{\mathbf{T}}^{-1}}$ and ${\mathbf{D}_{1}}=diag(\partial_{t_1},\partial_{t_1}+{{c}_{1\gamma}}{{\partial }_{\gamma1}},...,\partial_{t_1}+{{c}_{18\gamma}}{{\partial }_{\gamma1}}$). ${{{\tilde{\mathbf{D}}}}_{1}}$ can be computed as
\begin{equation}
\mathbf{T}{\mathbf{D}_{1}}{\mathbf{T}^{-1}}=\mathbf{T}({{\partial }_{t_1}}\mathbf{I}+{\mathbf{C}_{x}}{{\partial }_{x1}}+{\mathbf{C}_{y}}{{\partial }_{y1}}+{\mathbf{C}_{z}}{{\partial }_{z1}}){\mathbf{T}^{-1}}={{\partial }_{t_1}}\mathbf{I}+{\mathbf{\hat{C}}_{k}}{{\partial }_{k1}},
\end{equation}
where ${\mathbf{\hat{C}}_{k}}=\mathbf{T}{\mathbf{C}_{k}}{\mathbf{T}^{-1}},\ k=x,y,z$, $\mathbf{C}_k$ is a diagonal matrix with the $k$ component of every discrete velocity $\bm{c}_{i}$ as the diagonal element. ${\mathbf{\hat{C}}_{k}}$ are computed as
{\tiny
\begin{equation}
\mathbf{\hat{C}}_x=\left( {\begin{array}{*{20}c}
0  & 0    & 0   & 1 & 0  &   0  &   0  &   0   & 0  & 0  & 0   &  0   &  0    & 0    & 0     & 0   & 0   & 0   & 0 \\
0  & 0    & 0   & \frac{21}{5}  & \frac{19}{5}  &0  & 0  & 0   & 0  & 0  & 0   &  0   &  0    & 0    & 0     & 0   & 0   & 0   &  0\\
0  & 0    &0    & 0    & 1   & 0  & 0 & 0 & 0  & 0  & 0  & 0   &  0   &  0    & 0    & 0     & 0   & 0   & 0\\
\frac{10}{19} &   \frac{1}{57}  & 0   & 0    & 0  & 0   & 0  & 0  & 0  & \frac{1}{3} &  0   &  0    & 0    & 0     & 0   & 0   & 0  & 0  & 0 \\
0 & \frac{4}{63}  & \frac{10}{63} & 0   & 0    & 0  & 0   & 0  & 0  & -\frac{2}{9}  & \frac{5}{9}  &0   &0  &0 &0   &0    &0   &0   &0\\
0 &  0   & 0   & 0  & 0  & 0 & 0 & 0  & 0   &0  &0 &  0 &0   &1  &0   &0  &0   &0   &0\\
0 &  0   & 0   & 0  & 0  & 0  & 0  &0  &0  &0   &0   &0  &0  &1  &0   &0  &0   &0   &0\\
0 &  0   & 0   & 0  & 0  & 0  & 0  &0  &0  &0   &0   &0  &0  &0  &0   &1  &0   &0   &0\\
0 &  0   & 0   & 0  & 0  & 0  & 0  &0  &0  &0   &0   &0  &0  &0  &0   &1  &0   &0   &0\\
0 &  0   & 0   & \frac{6}{5}  & -\frac{1}{5}  & 0  & 0  &0  &0  &0   &0   &0  &0  &0  &0   &0  &0   &0   &0\\
0 &  0   & 0   & 0  & 1  & 0  & 0  &0  &0  &0   &0   &0  &0  &0  &0   &0  &0   &0   &0\\
0 &  0   & 0   & 0  & 0  & 0  & 0  &0  &0  &0   &0   &0  &0  &0  &0   &0  &1   &0   &0\\
0 &  0   & 0   & 0  & 0  & 0  & 0  &0  &0  &0   &0   &0  &0  &0  &0   &0  &1   &0   &0\\
0 &  0   & 0   & 0  & 0  & \frac{2}{5}  & \frac{1}{10}  &0  &0  &0   &0   &0  &0  &0  &0   &0  &0   &-\frac{1}{2}   &0\\
0 &  0   & 0   & 0  & 0  & 0  & 0  &0  &0  &0   &0   &0  &0  &0  &0   &0  &0   &0   &0\\
0 &  0   & 0   & 0  & 0  &  0 &  0 &\frac{2}{5}  &\frac{1}{10}  &0   &0   &0  &0  &0  &0   &0  &0   &0   &\frac{1}{2}\\
0 &  0   & 0   & 0  & 0  & 0  & 0  &0  &0  &0   &0   &\frac{2}{3}  &\frac{1}{3}  &0  &0   &0  &0   &0   &0\\
0 &  0   & 0   & 0  & 0  & 0  & 0  &0  &0  &0   &0   &0  &0  &-1  &0   &0  &0   &0   &0\\
0 &  0   & 0   & 0  & 0  & 0  & 0  &0  &0  &0   &0   &0  &0  &0  &0   &1  &0   &0   &0\\
 \end{array} } \right).
\end{equation}}

{\tiny
\begin{equation}
\mathbf{\hat{C}}_y=\left( {\begin{array}{*{20}c}
0  & 0    & 0   & 0 & 0  &   1  &   0  &   0   & 0  & 0  & 0   &  0   &  0    & 0    & 0     & 0   & 0   & 0   & 0 \\
0  & 0    & 0   & 0 & 0   &\frac{21}{5}  & \frac{19}{5}  & 0   & 0  & 0  & 0   &  0   &  0    & 0    & 0     & 0   & 0   & 0   &  0\\
0  & 0    &0    & 0    & 0   & 0  & 1 & 0 & 0  & 0  & 0  & 0   &  0   &  0    & 0    & 0     & 0   & 0   & 0\\
0 &   0  & 0   & 0    & 0  & 0   & 0  & 0  & 0  & 0 &  0   &  0    & 0    & 1     & 0   & 0   & 0  & 0  & 0 \\
0 &   0  & 0   & 0    & 0  & 0   & 0  & 0  & 0  & 0 &  0   &  0    & 0    & 1     & 0   & 0   & 0  & 0  & 0 \\
\frac{10}{19} &  \frac{1}{57}   & 0   & 0  & 0  & 0 & 0 & 0  & 0   &- \frac{1}{6} &0 &  \frac{1}{2} &0   &0  &0   &0  &0   &0   &0\\
0 &  \frac{4}{63}   & \frac{10}{63}   & 0  & 0  & 0  & 0  &0  &0  &\frac{1}{9}   &-\frac{5}{18}   &-\frac{1}{3}  &\frac{5}{6}  &0  &0   &0  &0   &0   &0\\
0 &  0   & 0   & 0  & 0  & 0  & 0  &0  &0  &0   &0   &0  &0  &0  &1   &0  &0   &0   &0\\
0 &  0   & 0   & 0  & 0  & 0  & 0  &0  &0  &0   &0   &0  &0  &0  &1   &0  &0   &0   &0\\
0 &  0   & 0   & 0  & 0  & -\frac{3}{5}  & \frac{1}{10}  &0  &0  &0   &0   &0  &0  &0  &0   &0  &0   &-\frac{3}{2}   &0\\
0 &  0   & 0   & 0  & 0  & 0     & -\frac{1}{2}  &0  &0  &0   &0   &0  &0  &0  &0   &0  &0   &- \frac{3}{2}   &0\\
0 &  0   & 0   & 0  & 0  & \frac{3}{5}  & -\frac{1}{10}  &0  &0  &0   &0   &0  &0  &0  &0   &0  &0   &-\frac{1}{2}   &0\\
0 &  0   & 0   & 0  & 0  & 0    & \frac{1}{2}  &0  &0  &0   &0   &0  &0  &0  &0   &0  &0   &-\frac{1}{2}   &0\\
0 &  0   & 0   & \frac{2}{5}  & \frac{1}{10}  &0  & 0  &0  &0  &0   &0   &0  &0  &0  &0   &0  &\frac{1}{2}   &0   &0\\
0 &  0   & 0   & 0  & 0  & 0  & 0  &\frac{2}{5}  &\frac{1}{10}  &0   &0   &0  &0  &0  &0   &0  &0   &0   &-\frac{1}{2}\\
0 &  0   & 0   & 0  & 0  &  0 &  0 &0    &0  &0   &0   &0  &0  &0  &0   &0  &0   &0   &0\\
0 &  0   & 0   & 0  & 0  &  0 &  0 &0    &0  &0   &0   &0  &0  &1  &0   &0  &0   &0   &0\\
0 &  0   & 0   & 0  & 0  &  0 &  0 &0    &0  &-\frac{1}{3}   &-\frac{1}{6}   &-\frac{1}{3}  &-\frac{1}{6}  &0  &0   &0  &0   &0   &0\\
0 &  0   & 0   & 0  & 0  &  0 &  0 &0    &0  &0   &0   &0  &0  &0  &-1   &0  &0   &0   &0\\
 \end{array} } \right).
\end{equation}}

{\tiny
\begin{equation}
\mathbf{\hat{C}}_z=\left( {\begin{array}{*{20}c}
0  & 0    & 0   & 0 & 0  &   0  &   0  &   1   & 0  & 0  & 0   &  0   &  0    & 0    & 0     & 0   & 0   & 0   & 0 \\
0  & 0    & 0   & 0 & 0  &   0  &   0  & \frac{21}{5}  & \frac{19}{5}  & 0  & 0   &  0   &  0    & 0    & 0     & 0   & 0   & 0   &  0\\
0  & 0    & 0   & 0 & 0  &   0  &   0  &   0   & 1  & 0  & 0   &  0   &  0    & 0    & 0     & 0   & 0   & 0   & 0\\
0 &   0  & 0   & 0    & 0  & 0   & 0  & 0  & 0  & 0 &  0   &  0    & 0    & 0     & 0   & 1   & 0  & 0  & 0 \\
0 &   0  & 0   & 0    & 0  & 0   & 0  & 0  & 0  & 0 &  0   &  0    & 0    & 0     & 0   & 1   & 0  & 0  & 0 \\
0 &   0  & 0   & 0    & 0  & 0   & 0  & 0  & 0  & 0 &  0   &  0    & 0    & 0     & 1   & 0   & 0  & 0  & 0\\
0 &   0  & 0   & 0    & 0  & 0   & 0  & 0  & 0  & 0 &  0   &  0    & 0    & 0     & 1   & 0   & 0  & 0  & 0\\
\frac{10}{19} & \frac{1}{57}   & 0   & 0    & 0  & 0   & 0  & 0  & 0  & -\frac{1}{6} &  0   &  -\frac{1}{2}  & 0   & 0   & 0   & 0   & 0  & 0  & 0\\
0 &  \frac{4}{63}   & \frac{10}{63}   & 0  & 0  & 0  & 0  &0  &0  & \frac{1}{9}   & -\frac{5}{18}   &\frac{1}{3}  &-\frac{5}{6}  &0  &0   &0  &0   &0   &0\\
0 &  0   & 0   & 0  & 0  & 0    & 0  &-\frac{3}{5}  &\frac{1}{10}  &0   &0   &0  &0  &0  &0   &0  &0   &0   &\frac{3}{2}\\
0 &  0   & 0   & 0  & 0  & 0    & 0  &0  &-\frac{1}{2}  &0   &0   &0  &0  &0  &0   &0  &0   &0   &\frac{3}{2}\\
0 &  0   & 0   & 0  & 0  & 0    & 0  &-\frac{3}{5}  &\frac{1}{10}  &0   &0   &0  &0  &0  &0   &0  &0   &0   &-\frac{1}{2}\\
0 &  0   & 0   & 0  & 0  & 0    & 0  &0     &-\frac{1}{2}  &0   &0   &0  &0  &0  &0   &0  &0   &0   &-\frac{1}{2}\\
0 &  0   & 0   & 0  & 0  &0  & 0  &0  &0  &0   &0   &0  &0  &0  &0   &0  &0   &0   &0\\
0 &  0   & 0   & 0  & 0  & \frac{2}{5}  & \frac{1}{10}  &0  &0  &0   &0   &0  &0  &0  &0   &0  &0   &\frac{1}{2}   &0\\
0 &  0   & 0   & \frac{2}{5}& \frac{1}{10} & 0  & 0  &0  &0  &0   &0   &0  &0  &0  &0   &0  &-\frac{1}{2}   &0   &0\\
0 &  0   & 0   & 0  & 0  &  0 &  0 &0    &0  &0   &0   &0  &0  &1  &0   &-1  &0   &0   &0\\
0 &  0   & 0   & 0  & 0  &  0 &  0 &0    &0  &0   &0   &0  &0  &0  &1   &0  &0   &0   &0\\
0 &  0   & 0   & 0  & 0  &  0 &  0 &0    &0  &\frac{1}{3} &\frac{1}{6} &-\frac{1}{3}  &-\frac{1}{6}  &0  &0   &0  &0   &0   &0\\
 \end{array} } \right).
\end{equation}}After calculating ${\mathbf{\hat{C}}_{k}}\bm{m}^{(eq)},k=x,y,z$ and substituting them into Eq. (\ref{difeps1orderm}), we have
\begin{equation}\label{solvem}\tiny
{\partial _{{t_1}}}\left[ {\begin{array}{*{20}{c}}
{\delta \rho}\\
{-11\delta \rho+\frac{19\textbf{\emph{j}} \cdot \textbf{\emph{j}}}{\rho_0}}\\
{\omega_{\epsilon}\delta \rho+\frac{\omega_{\epsilon j}\textbf{\emph{j}} \cdot \textbf{\emph{j}}}{\rho_0}}\\
{j_x}\\
{-2j_x/3}\\
{j_y}\\
{-2j_y/3}\\
{j_z}\\
{-2j_z/3}\\
{(2j_x^2-j_y^2-j_z^2)/\rho_0}\\
{\omega_{xx}(2j_x^2-j_y^2-j_z^2)/\rho_0}\\
{(j_y^2-j_z^2)/\rho_0}\\
{\omega_{xx}(j_y^2-j_z^2)/\rho_0}\\
{j_xj_y/\rho_0}\\
{j_yj_z/\rho_0}\\
{j_xj_z/\rho_0}\\
{0}\\
{0}\\
{0}
\end{array}} \right] + {\partial _{x1}}\left[ {\begin{array}{*{20}{c}}
{{j_x}}\\
{ 5{j_x}/3}\\
{-2{j_x}/3}\\
{j_x^2/\rho_0 + \delta \rho/3}\\
A\\
{{j_x}{j_y}/\rho_0}\\
{{j_x}{j_y}/\rho_0}\\
{{j_x}{j_z}/\rho_0}\\
{{j_x}{j_z}/\rho_0}\\
{4j_x/3}\\
{-2j_x/3}\\
0\\
{0}\\
{j_y/3}\\
{0}\\
{j_z/3}\\
{\frac{(j_y^2-j_z^2)(2+\omega_{xx})}{3\rho_0}}\\
{-j_xj_y/\rho_0}\\
{j_xj_z/\rho_0}
\end{array}} \right] + {\partial _{y1}}\left[ {\begin{array}{*{20}{c}}
{{j_y}}\\
{5{j_y}/3}\\
{-2{j_y}/3}\\
{{j_x}{j_y}/\rho_0}\\
{{j_x}{j_y}/\rho_0}\\
{j_y^2/\rho_0 + \delta \rho/3}\\
{B}\\
{j_yj_z/\rho_0}\\
{j_yj_z/\rho_0}\\
{-2j_y/3}\\
{j_y/3}\\
{2j_y/3}\\
{-j_y/3}\\
{j_x/3}\\
{j_z/3}\\
{0}\\
{j_xj_y/\rho_0}\\
{\frac{(j_z^2 - j_x^2)(2+\omega_{xx})}{3\rho_0}}\\
{-j_yj_z/\rho_0}
\end{array}} \right] + {\partial _{z1}}\left[ {\begin{array}{*{20}{c}}
{{j_z}}\\
{ 5{j_z}/3}\\
{-2{j_z}/3}\\
{{j_x}{j_z}/\rho_0}\\
{{j_x}{j_z}/\rho_0}\\
{{j_y}{j_z}/\rho_0}\\
{{j_y}{j_z}/\rho_0}\\
{j_z^2/\rho_0+\delta \rho/3}\\
{C}\\
{-2{j_z}/3}\\
{j_z/3}\\
{-2{j_z}/3}\\
{j_z/3}\\
{0}\\
{j_y/3}\\
{j_x/3}\\
{-j_xj_z/\rho_0}\\
{j_yj_z/\rho_0}\\
{\frac{(j_x^2-j_y^2)(2+\omega_{xx})}{3\rho_0}}
\end{array}} \right] =  - \left[ {\begin{array}{*{20}{c}}
0\\
{{s_1^{'}}{m^{(1)}_{1}}}\\
{{s_2^{'}}{m^{(1)}_{2}}}\\
{0}\\
{{s_4^{'}}{m^{(1)}_{4}}}\\
{0}\\
{{s_6^{'}}{m^{(1)}_{6}}}\\
{0}\\
{{s_8^{'}}{m^{(1)}_{8}}}\\
{{s_9^{'}}{m^{(1)}_{9}}}\\
{{s_{10}^{'}}{m^{(1)}_{10}}}\\
{{s_{11}^{'}}{m^{(1)}_{11}}}\\
{{s_{12}^{'}}{m^{(1)}_{12}}}\\
{{s_{13}^{'}}{m^{(1)}_{13}}}\\
{{s_{14}^{'}}{m^{(1)}_{14}}}\\
{{s_{15}^{'}}{m^{(1)}_{15}}}\\
{{s_{16}^{'}}{m^{(1)}_{16}}}\\
{{s_{17}^{'}}{m^{(1)}_{17}}}\\
{{s_{18}^{'}}{m^{(1)}_{18}}}\\
\end{array}} \right]
\end{equation}
where $s'_i=s_i/\delta t$ and
\begin{subequations}
\begin{equation}
\footnotesize
A={(76+10\omega_{\epsilon j})\textbf{\emph{j}} \cdot \textbf{\emph{j}}/(63\rho_0)+(2-5\omega_{xx})(-2j_x^2+j_y^2+j_z^2)/(9\rho_0)+2(5\omega_{\epsilon} - 22)\delta \rho/63},
\end{equation}
\begin{equation}
\footnotesize
B={(76+10\omega_{\epsilon j})\textbf{\emph{j}} \cdot \textbf{\emph{j}}/(63\rho_0)+(2-5\omega_{xx})(j_x^2-2j_y^2+j_z^2)/(9\rho_0)+2(5\omega_{\epsilon}-22)\delta \rho/63},
\end{equation}
\begin{equation}
\footnotesize
C={(76+10\omega_{\epsilon j})\textbf{\emph{j}} \cdot \textbf{\emph{j}}/(63\rho_0)+(2-5\omega_{xx})(j_x^2+j_y^2-2j_z^2)/(9\rho_0)+2(5\omega_{\epsilon}-22)\delta \rho/63}.
\end{equation}
\end{subequations}

From Eq. (\ref{solvem}), we have
\begin{subequations}
\label{sm1sm15}
\begin{equation}
\label{sm1}
{{\partial }_{t_1}}(-11\delta \rho +19\bm{j} \cdot \bm{j}/{{\rho }_{0}})+\frac{5}{3}\nabla_1 \cdot \bm{j}=- {{{s}'}_{1}}{{m}^{(1)}}
\end{equation}
\begin{equation}
\label{sm9}
\frac{1}{{{\rho }_{0}}}{{\partial }_{t_1}}\left( 2j_{x}^{2}-j_{y}^{2}-j_{z}^{2} \right)+\frac{2}{3}(2{{\partial }_{x1}}{{j}_{x}}-{{\partial }_{y1}}{{j}_{y}}-{{\partial }_{z1}}{{j}_{z}})= - {{{{s}'}}_{9}}{{m}_9^{(1)}}
\end{equation}
\begin{equation}
\label{sm11}
\frac{1}{{{\rho }_{0}}}{{\partial }_{t_1}}\left( j_{y}^{2}-j_{z}^{2} \right)+\frac{2}{3}({{\partial }_{y1}}{{j}_{y}}-{{\partial }_{z1}}{{j}_{z}})=- {{{{s}'}}_{11}}{{m}_{11}^{(1)}}
\end{equation}
\begin{equation}
\label{sm13}
\frac{1}{{{\rho }_{0}}}{{\partial }_{t_1}}\left( {{j}_{x}}{{j}_{y}} \right)+\frac{1}{3}({{\partial }_{x1}}{{j}_{y}}+{{\partial }_{y1}}{{j}_{x}})=- {{{{s}'}}_{13}}{{m}_{13}^{(1)}}
\end{equation}
\begin{equation}
\label{sm14}
\frac{1}{{{\rho }_{0}}}{{\partial }_{t_1}}\left( {{j}_{y}}{{j}_{z}} \right)+\frac{1}{3}({{\partial }_{y1}}{{j}_{z}}+{{\partial }_{z1}}{{j}_{y}})=- {{{{s}'}}_{14}}{{m}_{14}^{(1)}}
\end{equation}
\begin{equation}
\label{sm15}
\frac{1}{{{\rho }_{0}}}{{\partial }_{t_1}}\left( {{j}_{x}}{{j}_{z}} \right)+\frac{1}{3}({{\partial }_{x1}}{{j}_{z}}+{{\partial }_{z1}}{{j}_{x}})=- {{{{s}'}}_{15}}{{m}_{(15)}^{(1)}}
\end{equation}
\end{subequations}From the rows corresponding to the conserved moments in Eq. (\ref{solvem}), we have
\begin{subequations}
\label{continuityandmomentum}
\begin{equation}
\label{conti}
 {{\partial }_{{{t}_{1}}}} \delta \rho =-{{\partial }_{{{\gamma }_{1}}}}\left( \rho_0 {{u}_{\gamma }} \right),
\end{equation}
\begin{equation}
{\partial _{{t_1}}}{\rho _0}{u_\alpha } + {\partial _{\gamma 1}}\left( {{\rho _0}{u_\gamma }{u_\alpha }} \right) =  - {\partial _{\alpha 1}}\left( {c_s^2\delta \rho } \right),
\end{equation}
\end{subequations}
where $c_s^2=1/3$. In fact, above equation is satisfied for both LBGK and MRT LB models, and can be applied to most lattices,
such as D2Q9, D3Q15 and D3Q19 lattices. From above equation, we have
\begin{equation}
\label{Ma3}
{\partial _{{t_1}}}{\rho _0}{u_\alpha }{u_\beta } =  - c_s^2({u_\alpha }{\nabla _{1\beta }}\delta \rho  +
{u_\beta }{\nabla _{1\alpha }}\delta \rho ) - {\nabla _{1\gamma }}{\rho _0}{u_\alpha }{u_\beta }{u_\gamma }  -
{u_\alpha }{u_\beta }{\nabla _{1\gamma }}{\rho _0}{u_\gamma },
\end{equation} indicating that ${\partial _{{t_1}}}{\rho _0}{u_\alpha }{u_\beta } \ (\alpha,\beta=x,y,z)$ is of order $O(Ma^3)$.
Therefore, the terms $\frac{1}{{{\rho }_{0}}}{{\partial }_{t_1}}\left( {{j}_{\alpha}}{{j}_{\beta}} \right),\alpha,\beta=x,y,z$
in Eqs. (\ref{sm1sm15}) can be omitted for the incompressible flows. After multiplying $\varepsilon$ to both sides of Eq. (\ref{sm1sm15}), we obtain
\begin{subequations}
\label{YuLuobasis}
\begin{equation}
\label{mneq1}
 \varepsilon m_{1}^{(1)}\approx -\frac{38{{\delta }_{t}}}{3{{s}_{1}}}\nabla \cdot \bm{j},
\end{equation}
\begin{equation}
\label{mneq3}
 \varepsilon m_{9}^{(1)}\approx -\frac{2{{\delta }_{t}}}{3{{s}_{9}}}(3{{\partial }_{x}}{{j}_{x}}-\nabla \cdot \bm{j}),
\ \ \ \  \varepsilon m_{11}^{(1)}\approx -\frac{2{{\delta }_{t}}}{3{{s}_{9}}}({{\partial }_{y}}{{j}_{y}}-{{\partial }_{z}}{{j}_{z}}),
\end{equation}
\begin{equation}
\label{mneq5}
 \varepsilon m_{13}^{(1)}\approx -\frac{{{\delta }_{t}}}{3{{s}_{9}}}({{\partial }_{x}}{{j}_{y}}+{{\partial }_{y}}{{j}_{x}}),
\ \ \ \  \varepsilon m_{14}^{(1)}\approx -\frac{{{\delta }_{t}}}{3{{s}_{9}}}({{\partial }_{y}}{{j}_{z}}+{{\partial }_{z}}{{j}_{y}}),
\end{equation}
\begin{equation}
\label{mneq6}
 \varepsilon m_{15}^{(1)}\approx -\frac{{{\delta }_{t}}}{3{{s}_{9}}}({{\partial }_{z}}{{j}_{x}}+{{\partial }_{x}}{{j}_{z}}).
\end{equation}
\end{subequations}
From the equations of $\varepsilon m_{1}^{(1)}$, $\varepsilon m_{9}^{(1)}$ and $\varepsilon m_{11}^{(1)}$, we can
compute the diagonal elements of strain rate tensor. And from the equations of $\varepsilon m_{13}^{(1)}$,
$\varepsilon m_{14}^{(1)}$ and $\varepsilon m_{15}^{(1)}$,
we can calculate the off-diagonal elements of strain rate tensor. The final formula is Eq. (\ref{LuoS}).

\subsection{Derivation of Chai formula in computing the strain rate tensor in MRT LB model}
Equilibrium distribution functions for He-Luo MRT LB model satisfy
\begin{subequations}
\label{cfeqmoment_}
\begin{equation}
\sum\nolimits_{i}{{\bm{c}_{i}}{\bm{c}_{i}}f_{i}^{(eq)}}=c_{s}^{2}\delta \rho \mathbf{I}+{{\rho }_{0}}\bm{u}\bm{u},
\end{equation}
\begin{equation}
\sum\nolimits_{i}{{\bm{c}_{i}}{\bm{c}_{i}}{\bm{c}_{i}}f_{i}^{(eq)}}=c_{s}^{2} \rho_0 \Delta \cdot  \bm{u}=c_{s}^{2} \rho_0 \left( {{\delta }_{\alpha \beta }}{{u}_{\gamma }}+{{\delta }_{\alpha \gamma }}{{u}_{\beta }}+{{\delta }_{\beta \gamma }}{{u}_{\alpha }} \right).
\end{equation}
\end{subequations}

From Eqs. (\ref{feps0order}), (\ref{feps1order}), (\ref{conti}), (\ref{cfeqmoment_}) and noting ${\mathbf{\Lambda}}={\mathbf{T}^{-1}}\hat\mathbf{\Lambda }\mathbf{T}$, we obtain
\begin{eqnarray}
\label{Sc}
\lefteqn {-\frac{1}{{{\delta }_{t}}}\sum\limits_{i}{{{c}_{i\alpha }}{{c}_{i\beta }}}{{({\mathbf{T}^{-1}}\hat{\mathbf{\Lambda} }\mathbf{T})}_{ij}}f_{j}^{(1)}=\sum\limits_{i}{{{c}_{i\alpha }}{{c}_{i\beta }}{{D}_{1i}}f_{i}^{(0)}}} & \nonumber \\
 &={{\partial }_{{{t}_{1}}}}\left( \delta \rho c_{s}^{2}{{\delta }_{\alpha \beta }}+\rho_0 {{u}_{\alpha }}{{u}_{\beta }} \right)+{{\partial }_{{{\gamma }_{1}}}}\left[ \rho_0 c_{s}^{2}\left( {{u}_{\alpha }}{{\delta }_{\beta \gamma }}+{{u}_{\beta }}{{\delta }_{\alpha \gamma }}+{{u}_{\gamma }}{{\delta }_{\alpha \beta }} \right) \right]\nonumber \\
 & = c_s^2{\rho _0}\left( {{\partial _{{\alpha _1}}}{u_\beta } + {\partial _{{\beta _1}}}{u_\alpha }} \right) + {\partial _{{t_1}}}{\rho _0}{u_\alpha }{u_\beta }
\end{eqnarray}
Omitting the terms $O(Ma^3)$ (see Eq. (\ref{Ma3})) and multiplying $\varepsilon$ to the both sides of Eq. (\ref{Sc}), we obtain
\begin{equation}
\label{last}
- \frac{1}{{{\delta _t}}}\sum\limits_i {{c_{i\alpha }}{c_{i\beta }}} {({{\bf{T}}^{ - 1}}\hat {\mathbf{\Lambda }}{\bf{T}})_{ij}}({f_j} - f_j^{(eq)}) = c_s^2{\rho _0}\left( {{\partial _\alpha }{u_\beta } + {\partial _\beta }{u_\alpha }} \right).
\end{equation} From above equation, we can easily deduce the Chai formula (Eq. (\ref{ChaiS})) used for computing the strain rate tensor in MRT LB model.

\section{Discussion on two formulas in computing the strain rate tensor in MRT LB model}
\subsection{The equivalence of two formulas in computing the strain rate tensor in MRT LB model}
\label{equivalence}
In view of the form of formulas, Chai formula (Eq. (\ref{ChaiS})) is very different from Yu formula (Eq. (\ref{LuoS})).
Yu formula is based on non-equilibrium moments, while Chai formula is based on non-equilibrium density distribution functions.
To see if these two formulas are equivalent, we denote Chai formula using non-equilibrium moments. The proof is as follows.

Obviously,  ${\bm{m}^{(neq)}}=\mathbf{T}\left| {{f}^{(neq)}} \right\rangle $ and $\varepsilon {\bm{m}^{(1)}} = {\bm{m}^{(neq)}}$ are commonly used. Then we calculate $\mathbf{M}={\mathbf{T}^{-1}}\bm{\hat{\Lambda }}{\bm{m}^{(1)}}$, where $\mathbf{M}$ is column vector.
After that, we compute ${{N}_{\alpha \beta }}=\sum\nolimits_{i}{{{c}_{i\alpha }}{{c}_{i\beta }}{{M}_{i}}}$, where ${{M}_{i}}$ is the $i$-th element of column vector $\mathbf{M}$. The computation results are
\begin{subequations}
\begin{equation}
{{N}_{xx}}=\frac{{{s}_{1}}m_{1}^{(1)}}{57}+\frac{{{s}_{9}}m_{9}^{(1)}}{3}
\end{equation}
\begin{equation}
{{N}_{yy}}=\frac{{{s}_{1}}m_{1}^{(1)}}{57}-\frac{{{s}_{9}}m_{9}^{(1)}}{6}+\frac{{{s}_{9}}m_{11}^{(1)}}{2}
\end{equation}
\begin{equation}
{{N}_{zz}}=\frac{{{s}_{1}}m_{1}^{(1)}}{57}-\frac{{{s}_{9}}m_{9}^{(1)}}{6}-\frac{{{s}_{9}}m_{11}^{(1)}}{2}
\end{equation}
\begin{equation}
{{N}_{xy,yz,xz}}={{s}_{9}}m_{13,14,15}^{(1)}.
\end{equation}
\end{subequations}
Finally, we calculate ${{S}_{\alpha \beta }}=-\varepsilon {{N}_{\alpha \beta }}/\left( 2{{\rho }_{0}}c_{s}^{2}{{\delta }_{t}} \right)$, where ${{S}_{\alpha \beta }}$ is the element of strain rate tensor. Surprisingly, the final formula obtained from above computation is the same with Yu formula (Eq. (\ref{LuoS})).

In Ref. \cite{Chai2}, the accuracy of Chai formula is proved to be second-order accurate in space.
However, the accuracy of Yu formula has not been studied in previous literature.
Because the equivalence of these two formulas is proved,
it can be concluded that Yu formula is also second-order accurate in space.

\subsection{The computational efficiency of two formulas in computing the strain rate tensor in MRT LB model}
To calculate the strain rate tensor with Chai formula, we must firstly calculate all the elements of column vector ${\bf{T}}\left| {{f^{(neq)}}} \right\rangle$, where $\bf{T}$ is a $19\times19$ transformation matrix, $\left| {{f^{(neq)}}} \right\rangle$ is a column vector and is composed of non-equilibrium distribution functions in nineteen discrete velocity directions. We then calculate ${{\bf{T}}^{ - 1}}\left( {{\bf{\hat \Lambda T}}\left| {{f^{({{neq}})}}} \right\rangle } \right)$, where $\bm{\hat \Lambda}$ is a diagonal matrix and ${{\bf{T}}^{ - 1}}$ is a $19\times19$ matrix. Next, we have to calculate $\sum\limits_{i = 1}^{18} {{c_{i\alpha }}{c_{i\beta }}{{\left[ {{{\bf{T}}^{ - 1}}\left( {\bm{\hat \Lambda} {\bf{T}}\left| {{f^{(neq)}}} \right\rangle } \right)} \right]}_i}}$, where $\alpha,\beta=x,y,z$. Finally, the computation results are divided by $-2\rho_0 c_{s}^{2}{{\delta }_{t}}$ to obtain the strain rate tensor as Eq. (\ref{ChaiS}).

Compared with calculating the strain rate tensor by Chai formula, calculating the strain rate tensor by Yu formula requires much less computation quantity. Through Yu formula, we firstly calculate only 6 elements of column vector ${\bf{T}}\left| {{f^{(neq)}}} \right\rangle$, which are 2nd, 10th, 12th, 14th, 15th and 16th elements of column vector ${\bf{T}}\left| {{f^{(neq)}}} \right\rangle$. The above elements are approximately equal to $\varepsilon m_1^{(1)}$, $\varepsilon m_9^{(1)}$, $\varepsilon m_{11}^{(1)}$, $\varepsilon m_{13}^{(1)}$, $\varepsilon m_{14}^{(1)}$, $\varepsilon m_{15}^{(1)}$. Then we do some arithmetic operations to obtain the strain rate tensor as Eq. (\ref{LuoS}). From quantitative analysis, it is known that 957 times of multiplication and 792 times of addition are needed for Chai formula to work out the strain rate tensor at one grid point while only 155 times of multiplication/division and 113 times of addition/subtraction are required for Yu formula. Therefore, to calculate the strain rate tensor in MRT LB model more efficiently, Yu formula is a better choice.

\subsection{The generality of two formulas in computing the strain rate tensor in MRT LB model}
Yu formula is based on Eq. (\ref{YuLuobasis}), which is originated from Eq. (\ref{solvem}) or Eq. (\ref{sm1sm15}). Obviously, Eq. (\ref{solvem}) and Eq. (\ref{sm1sm15}) are only satisfied for D3Q19 lattice. Therefore, Yu formula can only be applied to D3Q19 MRT LB model to compute the strain rate tensor.

Compared with Yu formula, Chai formula is more general to compute the strain rate tensor for MRT LB model. Chai formula is derived mainly from Eq. (\ref{Sc}), which is stemmed from Eqs. (\ref{difepsorder}), (\ref{continuityandmomentum}) and (\ref{cfeqmoment_}). Interestingly, Eqs. (\ref{difepsorder}), (\ref{continuityandmomentum}) and (\ref{cfeqmoment_}) are satisfied for most lattices in the MRT LB model. Thus, Chai formula can be used to calculate the strain rate tensor not only for D3Q19 lattice, but also for other lattices, such as D2Q9 and D3Q15 lattices in the MRT LB model. From above analysis, it can be inferred that, if we want to compute the strain rate tensor for MRT LB model in a unified framework, Chai formula is a better choice.

\subsection{Two ways to obtain Yu formula for other lattices in MRT LB model}
Because Yu formula is more efficient in computing the strain rate tensor in MRT LB model, and we do not always use D3Q19 lattice, it is worthy to deduce the Yu formula for other lattices, such as the commonly used D2Q9 and D3Q15 lattices. For D$d$Q$q$ lattice, the derivation of Yu formula requires the computation of $d$ matrices ${\mathbf{\hat{C}}_{k}}=\mathbf{T}{\mathbf{C}_{k}}{\mathbf{T}^{-1}}$ and $d$ vectors ${\mathbf{\hat{C}}_{k}}\bm{m}^{(eq)}$, where $\mathbf{T}$, ${\mathbf{C}_{k}}$ and ${\mathbf{T}^{-1}}$ are $q\times q$ matrices. For example, the derivation of Yu formula for D3Q15 lattice needs to compute ${\mathbf{\hat{C}}_{k}}=\mathbf{T}{\mathbf{C}_{k}}{\mathbf{T}^{-1}}$ and $\ {\mathbf{\hat{C}}_{k}}\bm{m}^{(eq)}\ (k=x,y,z)$, where $\mathbf{T}$, ${\mathbf{C}_{k}}$ and ${\mathbf{T}^{-1}}$ are $15\times 15$ matrices. After above calculation, we can obtain similar equations to Eq. (\ref{solvem}), Eq. (\ref{sm1sm15}) and Eq. (\ref{YuLuobasis}). Finally, we do some tedious arithmetic operations to deduce the final Yu formula. It should be noted that above derivation can not be conducted only by symbol calculation but needs lots of manual work, which make the derivation very time-consuming.

Compared with above procedure, the derivation of Yu formula from Chai formula is more convenient. The derivation steps are the same with that in section \ref{equivalence}. Firstly, $\varepsilon {\bm{m}^{(1)}}$ is used to instead of $\mathbf{T}\left| {{f}^{(neq)}} \right\rangle $ in Eq.(\ref{ChaiS}). Then we compute $\mathbf{M}={\mathbf{T}^{-1}}\bm{\hat{\Lambda }}{\bm{m}^{(1)}}$ and ${{N}_{\alpha \beta }}=\sum\nolimits_{i}{{{c}_{i\alpha }}{{c}_{i\beta }}{{M}_{i}}}$, where ${{M}_{i}}$ is the $i$-th element of column vector $\mathbf{M}$. Finally, we compute ${{S}_{\alpha \beta }}=-\varepsilon {{N}_{\alpha \beta }}/\left( 2{{\rho }_{0}}c_{s}^{2}{{\delta }_{t}} \right)$ and thus obtain the Yu type of computational formula for stain rate tensor. Fortunately, above derivation way can be conducted completely by symbol calculation, which makes this way more convenient to deduce Yu formula.

\section{Conclusion}
\label{conclusions}
In this paper, we take He-Luo D3Q19 MRT LB model for an example to study the only two existing computational formulas for strain rate tensor in MRT LB model. These two formulas are named Yu formula and Chai formula in this paper. In view of the expression forms of these two formulas, it seems difficult to establish their correlation. However, through the theoretical analysis and symbol computation, it is found that these two formulas are actually equal to each other. In addition, the advantages and disadvantages of these two formulas are found. Yu formula is deduced through much more matrix computation and arithmetic operations, thus this formula is more specific to one type of MRT LB model and is more efficient in computing the strain rate tensor compared with Chai formula. In the other hand, Chai formula is deduced from some basic equations for most MRT LB models, so Chai formula can be applied to more lattice patterns of MRT LB models than Yu formula. Finally, it is found that to deduce the Yu type of formulas for other lattices, such as D2Q9 lattice and D3Q15 lattice, Chai formula can be used and this way is more convenient than the way proposed by Yu et al. All in all, in this paper it is found that the only two formulas in computing the strain rate tensor for MRT LB model are equal to each other but have their own advantages. To compute the strain rate tensor more efficiently, Yu formula is recommended. To compute the strain rate tensor for different lattice patterns of MRT LB models in a unified framework, Chai formula is suggested.

\section*{Acknowledgment}
This work is supported by the National Natural Science Foundation of China (Grant No.11502124, Grant No.11302073), the Natural Science Foundation of Zhejiang Province (Grant No.LQ16A020001), the Scientific Research Fund of Zhejiang Provincial Education Department (Grant No.Y201533808), Natural Science Foundation of Ningbo (Grant No. 2016A610075) and is sponsored by K.C. Wong Magna Fund in Ningbo University.

\end{document}